\documentclass{article}

\usepackage{arxiv}

\usepackage[utf8]{inputenc} % allow utf-8 input
\usepackage[T1]{fontenc}    % use 8-bit T1 fonts
\usepackage{hyperref}       % hyperlinks
\usepackage{url}            % simple URL typesetting
\usepackage{booktabs}       % professional-quality tables
\usepackage{amsfonts}       % blackboard math symbols
\usepackage{nicefrac}       % compact symbols for 1/2, etc.
\usepackage{microtype}      % microtypography
\usepackage{lipsum}
\usepackage{graphicx}
\graphicspath{ {./images/} }
\usepackage{times}
\usepackage{latexsym}

\usepackage[T1]{fontenc}
\usepackage[utf8]{inputenc}

\usepackage{microtype}

\usepackage{graphicx}
\usepackage{amsmath,amssymb,amsfonts}
\usepackage{algorithmic}
\usepackage{graphicx}
\usepackage{textcomp}
\usepackage{xcolor}
\usepackage{times}
\usepackage{latexsym}
\usepackage{booktabs}
\usepackage{subcaption}
\usepackage{array}
\usepackage{float}
\usepackage{hyperref}
\usepackage[htt]{hyphenat}

\title{Free and Customizable Code Documentation with LLMs: A Fine-Tuning Approach}

\author{
 Sayak Chakrabarty \\
  Department of Computer Science and Engineering\\
  Northwestern University\\
  Evanston, IL 60208 \\
  \texttt{sayakchakrabarty2025@u.northwestern.edu} \\
   \And
 Souradip Pal \\
  School of Electrical \& Computer Engineering\\
  Purdue University\\
  West Lafayette, IN 47906 \\
  \texttt{pal43@purdue.edu} \\
  }

\begin{document}
\maketitle
\begin{abstract}
Automated documentation of programming source code is a challenging task with significant practical and scientific implications for the developer community. We present a large language model (LLM)-based application that developers can use as a support tool to generate basic documentation for any publicly available repository. Over the last decade, several papers have been written on generating documentation for source code using neural network architectures. With the recent advancements in LLM technology, some open-source applications have been developed to address this problem. However, these applications typically rely on the OpenAI APIs, which incur substantial financial costs, particularly for large repositories.
Moreover, none of these open-source applications offer a fine-tuned model or features to enable users to fine-tune. Additionally, finding suitable data for fine-tuning is often challenging. Our application addresses these issues which is available at \url{https://pypi.org/project/readme-ready/}.
\end{abstract}


\section{Introduction}
The integration of natural and programming languages is a research area that addresses tasks such as automatic documentation of source code, code generation from natural language descriptions, and searching for code using natural language queries. These tasks are highly practical, as they can significantly enhance programmer efficiency, and they are scientifically intriguing due to their complexity and the proposed relationships between natural language, computation, and reasoning \cite{chomsky1956three,miller2003cognitive,graves2014neural}.

\begin{figure*}[h!]
    \centering
    \includegraphics[width=0.8\textwidth]{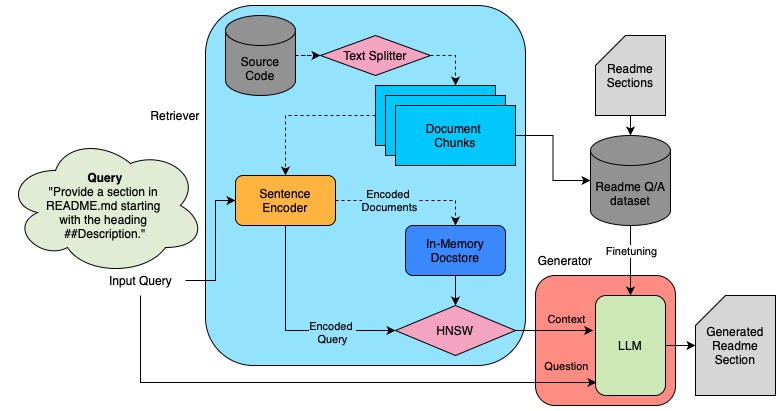} % Change image name and path as needed
    \caption{Input to Output Workflow showing the Retrieval and Generator modules. The retrieval module uses HNSW algorithm to create a context for the prompt to the Language model for text generation.}
    \label{fig:rag_workflow}
\end{figure*}

\subsection{Background and Related Work}
\vspace{-1mm}
Significant progress in machine learning\cite{chakrabarty2023single,datta2024consistency,bolonkin2024judicial} and other challenging natural language processing tasks has been achieved using neural networks, such as sequence-to-sequence transducers \cite{bahdanau2014neural}. Neural networks require training on extensive and diverse datasets\cite{zhang2022new} for effective generalization. These methods have been applied to code documentation \cite{iyer2016summarizing,barone2017parallel} and code generation \cite{ling2016latent,yin2017syntactic}, often using small or domain-specific datasets, sometimes confined to single software projects. Datasets like DJANGO and Project Euler \cite{oda2015learning} were developed by human annotators, ensuring accuracy but at a high cost and resulting in limited data sizes. Others, such as those referenced in \cite{allamanis2013mining,iyer2016summarizing} and IFTTT \cite{quirk2015language}, are larger but contain more noise.

Recently, large language models (LLMs) have become increasingly significant, demonstrating human-like abilities across various fields \cite{radford2019language,brown2020language,ouyang2022training}. LLMs typically employ transformer architecture variants and are trained on massive data volumes to detect patterns \cite{vaswani2017attention}.

In this paper, we focus on the automated documentation of programming source code, which is a challenging task with significant practical and scientific implications for the developer community. There are hundreds of publicly available repositories that lack basic documentation on aspects such as ``What does the repository do?", ``How to install the package?", ``How to run the code?", and ``How to contribute and modify any part of the repository?". None of the related research discussed handles these tasks using LLMs; instead, they design neural networks to address them.

We present an LLM-based application that developers can use as a support tool to generate basic documentation for any code repository. Some open-source applications have been developed to address this issue, to name a few
\begin{itemize}
\setlength{\itemsep}{0pt}
\item \texttt{AutoDoc-ChatGPT} \cite{autodoc-chatgpt}
\item \texttt{AutoDoc} \cite{context-labs-autodoc}
\item \texttt{Auto-GitHub-Docs-Generator} \cite{microsoft-auto-github-docs-generator}
\end{itemize}
However, these applications suffer from two major issues. Firstly, all of them are built on top of the OpenAI APIs, requiring users to have an OpenAI API key and incurring a cost with each API request. Generating documentation for a large repository could result in costs reaching hundreds of dollars. Our application allows users to choose among OpenAI's GPT, Meta's Llama2, and Google's Gemma models. Notably, apart from the first, the other models are open-source and incur no charges, allowing documentation to be generated for free.

Secondly, none of the existing open-source applications provide a fine-tuned model or an option for users to fine-tune their model on specific datasets. Our application offers a fine-tuning option using QLoRA, which can be trained on user's own dataset. It is important to note that using this feature requires access to powerful GPU clusters. Some existing applications provide a command-line tool for interacting with the entire repository, allowing users to ask specific questions about the repository but not generating a README file. We summarize our contributions below:
\subsection{Our Contributions}
\begin{itemize}
\item We propose a software package\footnote{https://github.com/souradipp76/ReadMeReady} that generates documentation for an open-source repository and serves as a documentation assistant for developers. This application creates a README file in Markdown format and allows generating a GitHub page from the documentation.
\item Instead of relying on the OpenAI APIs for handling calls, our model allows users to choose from various open-source models before generating the README, thereby avoiding the costs associated with existing applications.
\item As an example for our application, we fine-tune one of the models using LoRA. We allow users to fine-tune any of the models on a dataset we have created. Users' fine-tuning results may surpass ours due to our limited GPU resources.
\end{itemize}

\section{Methodology}
The application prompts the user to enter the project's name, and GitHub URL, and select the desired model from the following options:
\begin{itemize}
\setlength{\itemsep}{0pt}
\item \texttt{gpt-3.5-turbo} \cite{gpt-3.5-turbo}
\item \texttt{gpt-4} \cite{gpt-4}
\item \texttt{gpt-4-32k} \cite{gpt-4-32k}
\item \texttt{TheBloke/Llama-2-7B-Chat-GPTQ} (quantized) \cite{llama-2-7b-chat-gptq}
\item \texttt{TheBloke/CodeLlama-7B-Instruct-GPTQ} (quantized) \cite{code-llama-7b-instruct-gptq}
\item \texttt{meta-llama/Llama-2-7b-chat-hf} \cite{llama-2-7b-chat-hf}
\item \texttt{meta-llama/CodeLlama-7b-Instruct-hf} \cite{code-llama-7b-instruct-hf}
\item \texttt{google/gemma-2b-it} \cite{gemma-2b-it}
\item \texttt{google/codegemma-2b-it} \cite{codegemma-2b-it}
\end{itemize}
Note that the first two options will incur a cost for each call, and users need to provide an OpenAI API key. For large projects, the cost can reach several hundred dollars. Detailed OpenAI pricing can be found at \url{https://openai.com/api/pricing/}.

\textbf{Document Retrieval:} Our application indexes the codebase through a depth-first traversal of all repository contents and utilizes an LLM to generate documentation. All files are converted into text, tokenized, and then chunked, with each chunk containing 1000 tokens. The application employs the \texttt{sentence-transformers/all-mpnet-base-v2} \cite{sentence-transformers-all-mpnet-base-v2} sentence encoder to convert each chunk into a 768-dimensional embedding vector, which is stored in an in-memory vector store. When a query is provided, it is converted into a similar vector using the same sentence encoder. The neighbor nearest to the query embedding vector is searched using KNN (k=4) from the vector store, utilizing cosine similarity as the distance metric. For the KNN search, we use the HNSWLib library, which implements an approximate nearest-neighbor search based on hierarchical navigable small-world graphs \cite{malkov2018efficient}. This methodology provides the relevant sections of the source code, aiding in answering the prompted question. The entire methodology for Retrieval Augmented Generation(RAG) and fine-tuning is illustrated in Figure~\ref{fig:rag_workflow}.

\textbf{Prompt Configuration:} Prompt engineering is accomplished using the Langchain API. For our purpose, a prompt template has been used as provided in Appendix \ref{appendix:prompt_template}. This template includes placeholders for questions, which users can edit and modify as needed. This flexibility allows the README to be generated according to the user's specific requirements. Our default README structure includes sections on description, requirements, installation, usage, contributing methods, and licensing, which align with standard documentation practices. The temperature for text generation is kept at the default value of 0.2. The current prompts are developer-focused and assume that the repository is code-centric.

\subsection{Fine Tuning}
\vspace{-1mm}
Parameter-efficient fine-tuning (PEFT) \cite{lester2021power} is a technique in natural language processing that enhances pre-trained language models for specific tasks by fine-tuning only a subset of their parameters. This method involves freezing most of the model's layers and adjusting only the last few, thus conserving computational resources and time. Several parameter-efficient fine-tuning (PEFT) methods exist, such as Adapters, LoRA \cite{hu2022lora}, etc. We chose to fine-tune with QLoRA \cite{dettmers2023qlora} due to its significant reduction in the number of trainable parameters while maintaining performance. Given our limited resources, QLoRA is highly efficient as it adapts models for specific tasks with minimal computational overhead.

In our work, we fine-tune only one model, \texttt{TheBloke/Llama-2-7B-Chat-GPTQ} \cite{llama-2-7b-chat-gptq}, which is a 4-bit quantized model with 1.13 billion parameters. It supports a maximum sequence length of 4096 tokens and requires 3.9 GB of memory. We utilized GPU clusters provided by Northwestern University for fine-tuning our model. The configuration used is $1 \times$ NVIDIA Tesla V100 with 16GB of GPU memory. With this resource, training on a large dataset (12,803 data points) takes more than 15 hours, while training on a small dataset (339 data points) takes approximately 30 minutes for 3 epochs.

These resources are substantially limited compared to typical LLM fine-tuning requirements. Due to these constraints, we could only train the model for 3 epochs on a small dataset. As a result, we have made fine-tuning an optional feature, giving users the choice to fine-tune the model using their own GPU resources.

\vspace{-1mm} 
\subsection{Data Collection}
\vspace{-1mm}

Approximately 200 repositories were scraped using the GitHub APIs, selected based on popularity and star count. We limit our scope to Python-based repositories; however, this approach is easily adaptable to multiple programming languages. In scenarios involving various programming languages, distinct datasets can be created for fine-tuning purposes. A CSV file was created with three features: questions, context, and answers. Questions were derived from README file headings and subheadings, identified by markdown signatures "\#" or "\#\#". Answers correspond to the text under these headings. In our case, data consent is not required as the data is collected by scraping publicly available GitHub repositories.

The entire source code from the repositories is concatenated into a single string and separated into document chunks of 1000 tokens employing LangChain's text-splitter. Using the \texttt{sentence-transformers/all-mpnet-base-v2} \cite{sentence-transformers-all-mpnet-base-v2} sentence encoder, these chunks were converted into 768-dimensional vectors. Each question is then converted into a 768-dimensional vector and subjected to a KNN (k=4) search using HNSW\cite{malkov2018efficient} to find the closest match from the entire set of document embeddings, stored as the context.

\begin{table*}[t]
    \begin{tabular}{c}
        \begin{subtable}[!h]{0.39\textwidth}
            \centering
            \begin{tabular}{|l|c|c|}
                \hline
                \textbf{Repository} & \textbf{W/O FT} & \textbf{With FT} \\
                \hline
                allennlp & 32.09 & 16.38 \\
                autojump & 25.29 & 18.73 \\
                numpy-ml & 16.61 & 19.02 \\
                Spleeter & 18.33 & 19.47 \\
                TouchPose & 17.04 & 8.05 \\
                \hline
            \end{tabular}
            \caption{BLEU Scores}\label{BLEU}
        \end{subtable} 
        \begin{subtable}[h]{0.6\textwidth}
            \centering
            \begin{tabular}{|l|c|c|c|c|c|c|}
                \hline
                \textbf{Repository} & \multicolumn{3}{|c|}{\textbf{W/O Finetuning}} & \multicolumn{3}{|c|}{\textbf{With Finetuning}} \\
                \cline{2-7}
                & \textbf{P} & \textbf{R} & \textbf{F1} & \textbf{P} & \textbf{R} & \textbf{F1} \\
                \hline
                allennlp & 0.904 & 0.8861 & 0.895 & 0.862 & 0.869 & 0.865 \\
                autojump & 0.907 & 0.86 & 0.883 & 0.846 & 0.87 & 0.858 \\
                numpy-ml & 0.89 & 0.881 & 0.885 & 0.854 & 0.846 & 0.85 \\
                Spleeter & 0.86 & 0.845 & 0.852 & 0.865 & 0.866 & 0.865 \\
                TouchPose & 0.87 & 0.841 & 0.856 & 0.831 & 0.809 & 0.82 \\
                \hline
            \end{tabular}
            \caption{BERT Scores}\label{BERT}
        \end{subtable}
    \end{tabular}
    \caption{BLEU and BERT Scores of \texttt{TheBloke/Llama-2-7B-Chat-GPTQ} model for the generated README file across 5 code repositories before and after finetuning. The BERT scores shows the Precision(P), Recall(R) and F1 values when the contents are compared with each repositories' original README file.}
\end{table*}

\begin{table*}[t]
    \begin{tabular}{c}
        \begin{subtable}[!h]{0.4\textwidth}
            \centering
            \begin{tabular}{|l|l|}
                \hline
                \textbf{Parameter} & \textbf{Value} \\
                \hline
                \texttt{r} & 2 \\
                \texttt{lora\_alpha} & 32 \\
                \texttt{lora\_dropout} & 0.05 \\
                \texttt{bias} & None \\
                \texttt{task\_type} & CAUSAL\_LM \\
                \hline
                \end{tabular}
                \caption{QLoRA Configuration}\label{LoRA-config}
        \end{subtable} 
        \begin{subtable}[h]{0.5\textwidth}
            \centering
            \begin{tabular}{|l|l|}
            \hline
            \textbf{Parameter} & \textbf{Value} \\
            \hline
            \texttt{per\_device\_train\_batch\_size} & 1 \\
            \texttt{gradient\_accumulation\_steps} & 1 \\
            \texttt{num\_train\_epochs} & 3 \\
            \texttt{learning\_rate} & 1e-4 \\
            \texttt{fp16} & True \\
            \texttt{optim} & paged\_adamw\_8bit \\
            \texttt{lr\_scheduler\_type} & cosine \\
            \texttt{warmup\_ratio} & 0.01 \\
            \hline
            \end{tabular}
            \caption{Training Hyper-parameters}\label{train-params}
        \end{subtable}
    \end{tabular}
    \caption{Hyper-parameters for finetuning \texttt{TheBloke/Llama-2-7B-Chat-GPTQ} model including the configuration parameters for performing QLoRA.}
\end{table*}
\textbf{Data Preprocessing:} Following the creation of the CSV file, we pre-process the data using regex patterns to clean the text. Since the context only capture source code, this eliminates the possibility of using offensive content. Regex is used to remove hashtags, email addresses, usernames, image URLs, and other personally identifiable information. Note that only repositories written entirely in English are used, with other languages filtered out. Prompt engineering in our source code ensures that the prompts are designed to avoid generating any personally identifiable data or offensive content.

\section{Experiments and Results}
Our contribution is an application, not a new language generation model, making it challenging to establish an experimental methodology. We conducted the finetuning experiment on a small dataset consisting of randomly selected 190 README files, which may not address our default documentation questions. For each README, we examine its sections and subsections, frame relevant questions, and use the answers generated by our tool for training. A few samples of the fine-tuning dataset can be found in Appendix \ref{appendix:dataset}. For evaluation, we selected the rest 10 repositories and compared the original answers with the autogenerated documentation (sample documentation shown in Appendix \ref{appendix:output}) using BLEU and BERT scores to assess our model's performance.

\subsection{Before Fine-tuning}
We conducted a series of experiments utilizing the \texttt{TheBloke/Llama-2-7B-Chat-GPTQ} model \cite{llama-2-7b-chat-gptq} to demonstrate the functionality and efficacy of our proposed pipeline. The accompanying codebase is designed to be flexible, allowing the user to easily switch between different large language models (LLMs) by simply modifying the configuration file. Given the characteristics of LLMs, models with a greater number of parameters are generally expected to deliver enhanced performance. However, we lack the GPU resources to run a non-quantized version. The BLEU and BERT scores for the \texttt{TheBloke/Llama-2-7B-Chat-GPTQ} model are reported in Table~\ref{BLEU} and Table~\ref{BERT} respectively, under the ``W/O FT" or ``W/O Finetuning" columns.

\subsection{After Fine-tuning}
We utilized the PEFT library from Hugging Face, which supports several Parameter Efficient Fine-Tuning (PEFT) methods. This approach is cost-effective for fine-tuning large language models (LLMs), particularly on lightweight hardware. The training configuration and hyperparameters are detailed in Table~\ref{LoRA-config} and Table~\ref{train-params} respectively. The results are reported in Table~\ref{BERT} and Table~\ref{BLEU}, under the "With FT" or "With Finetuning" columns. it is observed that BLEU scores range from 15 to 30, averaging 20, indicating that the generated text is understandable but requires substantial editing to be acceptable. Conversely, BERT scores reveal a high semantic similarity to the original README content, with an average F1 score of ~85\%. The slightly lower scores for fine-tuned models compared to their original counterparts can be attributed to heavy quantization and the use of very low-rank configuration in QLoRA to manage memory constraints, leading to a noticeable reduction in quality. The effectiveness of the tool in generating accurate documentation relies significantly not only on the fine-tuned model for text generation but also on the comprehensiveness of the code context included in the prompt. Consequently, when relevant context is captured by the HNSW algorithm using text embeddings, the generated outputs remain satisfactory even when employing quantized fine-tuned models. However, the performance of these models tends to plateau if sufficient contextual information is not adequately captured.

\section{Conclusion}
Our application addresses the critical need for generating documentation for code repositories by utilizing multiple LLM models and allowing users to fine-tune these models using LoRa on their own GPUs. While our approach is not designed to surpass state-of-the-art benchmarks, its significance lies in the application of NLP techniques to solve a pressing issue faced by the developer community. The tool provides initial documentation suggestions based on the source code, assisting developers in initiating the documentation process and enabling them to modify the generated README files to meet their specific requirements, thereby reducing manual effort. Additionally, the generated README files can be seamlessly converted into PyPI-compliant standard documentation websites using tools such as MkDocs or Sphinx. This application can also be adapted as a plugin for integration with code editors like Visual Studio Code, thus enhancing the development workflow by minimizing the need for manual documentation creation.

\begin{table*}[t]
    \footnotesize
    \centering
    \begin{tabular}{|p{0.11\linewidth}|p{.13\linewidth}|p{.1\linewidth}|p{.35\linewidth}|p{.25\linewidth}|}
        \hline
        \textbf{project\_name} & \textbf{repository\_url} & \textbf{heading} & \textbf{context} & \textbf{answer}\\
        \hline
        magenta & magenta/magenta & Getting Started 
        & \texttt{Number of parameter servers. If 0, parameters ' 'are handled locally by the worker.') flags.DEFINE\_integer ('save\_summaries\_secs', 15, 'Frequency at which summaries are saved, in seconds.') ...flags.DEFINE\_string ('style\_dataset\_file', None, 'Style dataset file.') flags.DEFINE\_string('train\_dir', None, 'Directory for checkpoints and summaries.') flags.DEFINE\_string ('inception\_v3\_checkpoint}
        & Take a look at our [colab notebooks](URL) for various models, including one on [getting started](URL). [Magenta.js](URL) is also a good resource for models and [demos](URL) that run in the browser. This and more, including [blog posts](URL) and [Ableton Live plugins](URL), can be found at [https://magenta.tensorflow.org] (URL).\\
        \hline
        allennlp & allenai/allennlp & Plugins 
        & \texttt{def discover\_plugins() -> Iterable[str]: """""" Returns an iterable of the plugins found. """""" plugins: Set[str] = set() ...  def discover\_file\_plugins (plugins\_filename: str = LOCAL\_PLUGINS\_FILENAME) -> Iterable[str]: """""" Returns an iterable of the plugins found, declared within a file whose path is `plugins\_filename`. """""" with open(plugins\_filename) as file\_: for module\_name in file\_.readlines(): module\_name = module\_name.strip() if module\_name: yield module\_name}
        & AllenNLP supports loading "plugins" dynamically. A plugin is just a Python package that provides custom registered classes or additional `allennlp` subcommands. There is ecosystem of open source plugins, some of which are maintained by the AllenNLP team here at AI2, and some of which are maintained by the broader community. ... To test that your plugins can be found and imported by AllenNLP, you can run the `allennlp test-install` command. Each discovered plugin will be logged to the terminal. For more information about plugins, see the [plugins API docs](URL). And for information on how to create a custom subcommand to distribute as a plugin, see the [subcommand API docs](URL).\\
        \hline
        python-fire & google/python-fire & Installation 
        & \texttt{`Programming Language :: Python', `Programming Language :: Python :: 3', `Programming Language :: Python :: 3.12', `Operating System :: OS Independent', `Operating System :: POSIX', `Operating System :: MacOS', `Operating System :: Unix', ], keywords=`command line interface cli python fire interactive bash tool', packages=[`fire', `fire.console'],packages=[`fire', `fire.console'], ... (`Darwin', `<hostname goes here>', `12.4.0', \# `Darwin Kernel Version 12.4.0: Wed May 1 17:57:12 PDT 2013; \# root:xnu-2050.24.15\~ 1/RELEASE\_X86\_64', `x86\_64', `i386') format\_string = `(Macintosh; {name} Mac OS X {version})'}
        & To install Python Fire with pip, run: `pip install fire` To install Python Fire with conda, run: `conda install fire -c conda-forge` To install Python Fire from source, first clone the repository and then run: `python setup.py install`\\
        \hline
    \end{tabular}
    \caption{Samples of input (project\_name, heading, context) and output (answers) for prompts used in fine-tuning \texttt{TheBloke/Llama-2-7B-Chat-GPTQ} model}\label{data}
\end{table*}

\bibliographystyle{unsrt}  
\bibliography{references}

\appendix

\section{Ethical Limitations}
Like any study, ours has its limitations. It is important to note that we propose our application as a developer support tool. While it assists developers in documenting unseen repositories or undocumented code bases and quickly understanding how to use them, developers should not blindly follow the instructions provided by the application. This caution is necessary because the application relies on pre-trained models like Meta's Llama2 and OpenAI's GPT, which are known to hallucinate and provide incorrect instructions with confidence. We have provided a fine-tuning feature and plan to launch the fine-tuned model, which is expected to hallucinate less than the raw model. Another limitation is that the application will not address more complex questions that developers might have; it will only provide answers to basic questions. However, developers can modify our source code to add more prompts to obtain specific answers. In such cases, fine-tuning must be performed again with a different dataset. Lastly, no AI assistants were used for this research or the writing of this paper.

The primary potential risk of this application is that it might misguide developers if they uncritically accept every instruction generated in the README file. Our experiments indicate that the generated README files produce close to correct content (as measured by various evaluation methods; see Table~\ref{BERT} and Table~\ref{BLEU}), but it is still crucial not to take every piece of text generated by the LLM backend at face value. The application should be regarded as a support tool for developers, rather than a definitive source of truth.

\section{Data Samples}\label{appendix:dataset}
Table \ref{data} shows samples of data that were used in fine-tuning the pretrained language models based on the prompt template. For our experiment, some of the inputs to the prompt were kept fixed but may be varied based on the use-case. Here, we fixed the \texttt{content\_type} as ``docs", \texttt{target\_audience} as ``smart developer" and did not pass any \texttt{additional\_instructions}. In Table \ref{data}, shortened versions of the context has been provided to give an idea on the contextual information obtained using the HNSW algorithm on the document chunks. 

\section{Prompt Template}\label{appendix:prompt_template}
\textbf{Instruction:} \texttt{You are an AI assistant for a software project called \{project\_name\}. You are trained on all the \{content\_type\} that makes up this project. The \{content\_type\} for the project is located at \{repository\_url\}. You are given a repository which might contain several modules and each module will contain a set of files. Look at the source code in the repository and you have to generate content for the section of a README.md file following the heading given below. If you use any hyperlinks, they should link back to the github repository shared with you. You should only use hyperlinks that are explicitly listed in the context. Do NOT make up a hyperlink that is not listed. Assume the reader is a \{target\_audience\} but is not deeply familiar with \{project\_name\}. Assume the reader knows nothing about how the project is structured or which folders/files do what and what functions are written in which files and what these functions do. If you don't know how to fill up the README.md file in one of its sections, leave that part blank. Don't try to make up any content. Do not include information that is not directly relevant to repository, even though the names of the functions might be common or is frequently used in several other places. Provide the answer in correct markdown format. \\
\{additional\_instructions\}}\\
\textbf{Question:} \texttt{Provide the README content for the section with heading "\{\{input\}\}" starting with \#\#\{\{heading\}\}.}\\
\textbf{Context}:
\texttt{\{\{context\}\}}\\
\textbf{Answer in Markdown:} \texttt{\{\{answer\}\}}

\section{Sample Autogenerated Readme}\label{appendix:output}
{\footnotesize
\begin{verbatim}
# python-fire
<pre><code>**Description**</code></pre>
<p>The <code>python-fire</code> library is a powerful tool for building web applications. 
It provides a wide range of features and functionalities that can help you build complex 
web applications quickly and efficiently. In this section, we will provide an overview of 
the main features and functionalities of <code>python-fire</code>.</p>

<pre><code>**Installation**
To install the latest version of `python-fire`, follow these steps:
Clone the repository from GitHub:
<p><code>bash
    git clone https://github.com/google/python-fire.git
</code></p>
<pre><code>Change into the cloned repository:</code>
</pre><p>```bash</p><pre><code>cd python-fire</code>
</pre><p>```</p><pre><code>Run the following command to install the dependencies:
</code></pre><p>```bash</p>
<pre><code>pip install -r requirements.txt</code></pre>
<p>```</p><pre><code>
This will install all the necessary dependencies for `python-fire`. Once the installation 
is complete, you can run `python-fire` by running the following command: 
</code></pre><p>```bash</p><pre>
<code>python fire</code></pre><p>```</p><pre><code>
This will launch `python-fire` in your default Python interpreter.

**How to contribute**
We'd love to accept your patches and contributions to this project. There are just a 
few small guidelines you need to follow. Before you begin making changes, state your 
intent to do so in an Issue. Then, fork the project. Make changes in your copy of the 
repository. Then open a pull request once your changes are ready. If this is your 
first contribution, sign the Contributor License Agreement. A discussion about your 
change will follow, and if accepted your contribution will be incorporated into the 
Python Fire codebase.
\end{verbatim}}
\end{document}